\documentclass[12pt]{article}
\usepackage{amsmath}
\usepackage{amssymb}
\usepackage{mathtext}
\usepackage{graphicx}

\begin{document}

\section*{DYNAMICS AND STATICS OF VORTICES ON A PLANE AND A SPHERE~---
I}\footnote{REGULAR AND CHAOTIC DYNAMICS V.~3, No.1, 1998\\Received October
1, 1997, revised manuscript received February 1, 1998}

\begin{centering}
A.\,V.\,BORISOV\\
Faculty of Mechanics and Mathematics,
Department of Theoretical Mechanics
Moscow State University\\
Vorob'ievy gory, 119899 Moscow, Russia\\
E-mail: borisov@uni.udm.ru\\
A.\,E.\,PAVLOV\\
Laboratory of Nonlinear Dynamics and Synergetics, Udmurt State University\\
Universitetskaya Str. 1, 426034. Izhevsk, Russia\\
E-mail: pavlov@uni.udm.ru\\
\end{centering}

\begin{abstract}
In the present paper a description of a problem of point vortices on a
plane and a sphere in the ``internal'' variables is discussed. The Hamiltonian
equations of motion of vortices on a plane are built on the Lie--Poisson
algebras, and in the case of vortices on a sphere on the quadratic Jacobi
algebras. The last ones are obtained by deformation of the corresponding
linear algebras. Some partial solutions of the systems of three and four
vortices are considered. Stationary and static vortex configurations are found.
\end{abstract}

\section{Dynamics of point vortices on a plane}

\subsection*{Introduction}

Let us consider the plane motion in boundless ideal liquid of
the~$N$~rectilinear filaments with the strengths~$\Gamma_i$, crossing the
plane in the points with coordinates~$(x_i,\,y_i)$. It was shown by
Kirchhoff~[1] that the motion equations of such system can be written
in the Hamiltonian form:
\begin{equation}
\label{e1_1}
\Gamma_i\dot x_i=\frac{\partial H}{\partial y_i}\mbox{, }\Gamma_i\dot y_i=-{\frac{\partial H}{\partial
x_i}}\mbox{, }1\leqslant i\leqslant n,
\end{equation}
with the Hamiltonian
\begin{equation}
\label{e1_2}
H=-\frac{1}{8\pi}\sum_{i,j=1}^N{'}\Gamma_i\Gamma_j\ln M_{ij},
\end{equation}

$$
M_{ij}\equiv(x_i-x_j)^2+(y_i-y_j)^2,
$$

The Poisson brackets are of the form:
\begin{equation}
\label{e1_3} \{f,\,g\}=\sum_{i=1}^N\frac{1}{\Gamma_i}\left(\frac{\partial
f}{\partial x_i} \frac{\partial g}{\partial y_i}-\frac{\partial
f}{\partial y_i}\frac{\partial g}{\partial x_i}\right).
\end{equation}

The system of equations~(\ref{e1_1}) has, beside of energy (\ref{e1_2}),
the first integrals due to the invariance of the Hamiltonian with respect
to translations and rotations of the coordinate system:
\begin{equation}
\label{e1_4}
Q=\sum_{i=1}^N\Gamma_ix_i\mbox{, } P=\sum_{i=1}^N\Gamma_iy_i\mbox{, }
I=\sum_{i=1}^N\Gamma_i(x_i^2+y_i^2).
\end{equation}

The set of integrals (\ref{e1_4}) is not in involution:
\begin{equation}
\label{e1_5}
\{Q,\,P\}=\sum_{i=1}^N\Gamma_i\mbox{, }\{P,\,I\}=-2Q\mbox{, }\{Q,\,I\}=2P.
\end{equation}

\subsection*{Representation in ``internal'' variables}


In this paper another representation of problem is given. Let us choose
the squares of mutual distances between the point vortices as new variables.
For systems of~$N$~vortices the number of mutual distances is equal
to~$N(N-1)/2$. Then, commuting~$M_{ij}$ and~$M_{lk}$ in the Poisson
structure~(\ref{e1_3}) and using the Heron formulas, we as a result obtain
some nonlinear Poisson brackets in the phase space determined by mutual
distances. However, if we add new variables, the oriented areas
of parallelograms spanned on triples of vortices with numbers~$i,\:j,\:k$:
\begin{equation}
\label{e1_6}
\triangle_{ijk}\equiv({\bf r}_j-{\bf r}_i)\wedge({\bf r}_k-{\bf
r}_i)\mbox{, }
{\bf r}_l=(x_l,\,y_l),
\end{equation}
then in the phase space of the
variables~$(M_{ij},\,\triangle_{ijk},\,i,\,j,\,k=1,\,\ldots\,,\,N)$
the Lie--Poisson brackets appear. The dimension of the phase is made
up of the number of mutual distances between vortices (the number of binomial
combinations from~$N$ on~$2$) and the number of triangles formed by the
triples of vortices (the number of binomial combinations of~$N$
things~${\bf r}$ at a time) (see~Fig.~1): $C_N^2+C_N^3=C_{N+1}^3$.

Calculation of the Poisson brackets between these variables using (\ref{e1_3})
leads to the following relations:
\begin{equation}
\label{e1_7}
\{M_{ij},\,M_{kl}\}=4\left(\frac{1}{\Gamma_i}\delta_{ik}-\frac{1}{\Gamma_j}\delta_{jk}\right)
\triangle_{ijl}+4\left(\frac{1}{\Gamma_i}\delta_{il}-\frac{1}{\Gamma_j}\delta_{jl}\right)\triangle_{ijk};
\end{equation}

\begin{equation}
\begin{split}
\label{e1_8}
\{M_{ij},\,\triangle_{klm}\} =\left(\frac{1}{\Gamma_i}\delta_{ik}-\frac{1}{\Gamma_j}\delta_{jk}
\right)(M_{li}-M_{im}+M_{mj}-M_{jl})+\\
+\left(\frac{1}{\Gamma_i}\delta_{il}-\frac{1}{\Gamma_j}\delta_{jl}\right)
(M_{mi}-M_{ik}+M_{kj}-M_{jm})+\\
+\left(\frac{1}{\Gamma_i}\delta_{im}-\frac{1}{\Gamma_j}\delta_{jm}\right)
(M_{ki}-M_{il}+M_{lj}-M_{jk});
\end{split}
\end{equation}

\begin{equation}
\begin{split}
\label{e1_9}
\{\triangle_{ijk},\,\triangle_{lmn}\} & = \frac{\delta_{il}}{\Gamma_i}(\triangle_{jkn}-\triangle_{jkm})+
\frac{\delta_{im}}{\Gamma_i}(\triangle_{jkl}-\triangle_{jkn})+\frac{\delta_{in}}{\Gamma_i}
(\triangle_{jkm}-\triangle_{jkl})+\\
+ & \frac{\delta_{jl}}{\Gamma_j}(\triangle_{ikm}-\triangle_{ikn})+
\frac{\delta_{jm}}{\Gamma_j}(\triangle_{ikn}-\triangle_{ikl})+
\frac{\delta_{jn}}{\Gamma_j}(\triangle_{ikl}-\triangle_{ikm})+\\
+ & \frac{\delta_{kl}}{\Gamma_k}(\triangle_{ijn}-\triangle_{ijm})+
\frac{\delta_{km}}{\Gamma_k}(\triangle_{ijl}-\triangle_{ijn})+
\frac{\delta_{kn}}{\Gamma_k}(\triangle_{ijm}-\triangle_{ijl}).
\end{split}
\end{equation}

Thus, in the~$C_{N+1}^3$-dimensional phase space of the variables~$(M,\,\triangle)$
the Lie--Poisson brackets are determined. The structure of this algebra is as
follows:
$$
\{M,\,M\}\subset\triangle\mbox{, }\{M,\triangle\}\subset M\mbox{,
}\{\triangle,\,\triangle\}\subset\triangle.
$$
Besides the subalgebra~$\triangle$, this algebra contains subalgebras of problems
of~$(N-1)$, $(N-2),\,\ldots$,~3~vortices.

The Poisson brackets are degenerated. The linear Casimir function
\begin{equation}
\label{e1_10}
\triangle=\sum_{i,\,j=1}^N\Gamma_i\Gamma_jM_{ij}
\end{equation}
is due to existence of the integral of angular momentum of vortex system.
The linear Casimir functions correspond to the trivial geometric relations
between the oriented areas of triangles. Let us write here, as an example,
one such relation for a
quadrangle~$\triangle_{ijk}+\triangle_{ikl}-\triangle_{lij}-\triangle_{ljk}=0$
(see~Fig.~1). It is easy to see that the annihilator of the brackets is
\begin{equation}
\label{e1_11}
F_{ijkl}=\triangle_{ijk}+\triangle_{ikl}-\triangle_{lij}-\triangle_{ljk}=0.
\end{equation}

The structure (\ref{e1_9}) has also the additional Casimir functions. They
came about from such relations as
\begin{equation}
\label{e1_12}
F_{ijk}=(2\triangle_{ijk})^2+M_{ij}^2+M_{jk}^2+M_{ik}^2-2(M_{ij}M_{jk}+
M_{ij}M_{ik}+M_{jk}M_{ik})=0
\end{equation}
derived from the Heron formula, relating the areas and lengthes of sides
of a triangle. In general even in the problem of four vortices the
functions (\ref{e1_12}) are not Casimir functions (they determine Casimir
functions of subalgebra of three vortices). The detailed analysis and
classification of Poisson brackets algebra (\ref{e1_9}) we plan to present
in the next part of our work (to be published).

If we express~$\triangle_{ijk}$ from (\ref{e1_12}) and substitute in the equations
of motion for the squares of mutual distances~$M_{ij}$, then we get the
equations of Laura~[2]. Having solved the equations of motion for the relative
positions of vortices, we are able to find, using the quadratures and the
initial conditions, the absolute coordinates of their positions on a plane
at any time~[3].

\subsection*{Partial cases of integrability of three and four vortices on
a plane}

The problems of three and four vortices are of particular interest. The
integrability of the problem of three vortices was indicated already
by Poincar\'e. The problem of four vortices is nonintegrable.

Under reduction of the ten-dimensional algebra, corresponding to the problem
of four vortices on the general level of six Casimir functions, the Poisson
structure becomes nondegenerate. So, according to the Darboux theorem,
there exist local canonical coordinates, in which the equations have the form
of a usual Hamiltonian system with two degrees of freedom. For integrability
of such system, besides the Hamiltonian, one more first integral is required,
which, in general case, does not exist.

Let us consider in detail a subalgebra (\ref{e1_7})--(\ref{e1_9}),
corresponding to the case of three vortices. In variables~$M$, $\triangle$ it has
the form:
\begin{equation}
\begin{split}
\label{e1_13}
\{M_1,\,M_2\}=-\frac{4}{\Gamma_3}\triangle_4\mbox{,
}\{M_3,\,M_1\}=-\frac{4}{\Gamma_2}\triangle_4\mbox{, }
\{M_2,\,M_3\}=-\frac{4}{\Gamma_1}\triangle_4,\\
\{M_1,\,\triangle_4\}=\left(\frac{1}{\Gamma_2}-\frac{1}{\Gamma_3}\right)M_1+
\left(\frac{1}{\Gamma_2}+\frac{1}{\Gamma_3}\right)(M_2-M_3),\\
\{M_2,\,\triangle_4\}=\left(\frac{1}{\Gamma_3}-\frac{1}{\Gamma_1}\right)M_2+
\left(\frac{1}{\Gamma_3}+\frac{1}{\Gamma_1}\right)(M_3-M_1),\\
\{M_3,\,\triangle_4\}=\left(\frac{1}{\Gamma_1}-\frac{1}{\Gamma_2}\right)M_3+
\left(\frac{1}{\Gamma_1}+\frac{1}{\Gamma_2}\right)(M_1-M_2).
\end{split}
\end{equation}

The rank of the algebra (\ref{e1_13}) is two. It corresponds on two-dimensional
symplectic leaf to the system with one degree of freedom. Depending on the
ratio of intensities, this algebra is a direct sum~$SO(3)\oplus\mathbb{R}$
or~$SO(1,\,2)\oplus\mathbb{R}$ (the case~$\Gamma_1+\Gamma_2+\Gamma_3=0$ is under
consideration). Analysis of this problem will be also given in our next
publication.

One can point out at an interesting analogy between the problem of three
vortices and the system of Lotka--Volterra, appearing in mathematical
biology~[4].

The Hamiltonian of three vortices
\begin{equation}
\label{e1_14}
H=-\frac{1}{8\pi}(\Gamma_2\Gamma_3\ln M_1+\Gamma_1\Gamma_3\ln M_2+\Gamma_1\Gamma_2\ln M_3)
\end{equation}
generates the phase flow:
\begin{equation}
\label{e1_15}
\hookrightarrow\dot M_1=\{M_1,\,H\}=\frac{\Gamma_1}{2\pi}
\left(\frac{1}{M_2}-\frac{1}{M_3}\right)\triangle_4,
\end{equation}
$$
\dot\triangle_4=\frac{1}{8\pi}\left(\frac{\Gamma_2+\Gamma_3}{M_1}(M_2-M_3)+
\frac{\Gamma_1+\Gamma_3}{M_2}(M_3-M_1)+\frac{\Gamma_1+\Gamma_2}{M_3}(M_1-M_2)\right).
$$

Expressing~$\triangle_4$ from (\ref{e1_12})

\begin{equation}
\label{e1_16}
\triangle_4=\pm\frac{1}{2}\sqrt{-M_1^2-M_2^2-M_3^2+2(M_1M_2+M_1M_3+M_2M_3)},
\end{equation}
we get the system:

\begin{equation}
\label{e1_17}
\hookrightarrow\dot M_1=\frac{\Gamma_1\triangle_4}{2\pi M_1M_2M_3}M_1(M_3-M_2).
\end{equation}

If we introduce the regularizing time~$\tau$:

\begin{equation}
\label{e1_18}
\frac{d\tau}{dt}=\frac{\triangle_4}{2\pi M_1M_2M_3},
\end{equation}
then we find that~$M_i$ satisfy the Lotka--Volterra system:
\begin{equation}
\label{e1_19}
\hookrightarrow\frac{d}{d\tau}M_1=\Gamma_1M_1(M_3-M_2).
\end{equation}

Under transition of the system of vortices to a collinear
configuration~$(\triangle_4=0)$ a change of time flow is possible, so one must
choose in (\ref{e1_16}) the sign plus. Hence, we have demonstrated a piecewise
trajectory isomorphism between two dynamical problems: the problem of three
vortices and the three-dimensional Lotka--Volterra system, which takes place
also in the problem of three vortices on a sphere, as it will be
shown in the following section.

Let us note a partial case of integrability of four-vortex problem~[5]:
\begin{equation}
\label{e1_20}
\sum_{i=1}^4\Gamma_i=0\mbox{, } Q=P=0.
\end{equation}
In this case, the system has sufficient number of integrals~$Q$, $P$, $I$,
in involution (see the algebra of integrals (\ref{e1_5})). Its corresponding
Casimir function is~$D$ on zero level: $D=0$ due to the relation between the
first integrals~[6]:
\begin{equation}
\label{e1_21}
D/2=\left(\sum_{i=1}^N\Gamma_i\right)I-Q^2-P^2.
\end{equation}

Additional invariant relations which are necessary for integrability in
variables~$M_{ij}$ one can find from the~$N(N-1)$ identities~[6]
\begin{equation}
\label{e1_22}
f_{ij} \equiv \frac{1}{2}\sum_{k=1}^N\Gamma_k(M_{jk}-M_{ik})=P(x_i-x_j)+Q(y_i-y_j),
\end{equation}
being valid under~$\sum_{i=1}^N\Gamma_i=0$.

Let us indicate, for example, a particular solution of the system of four
vortices if a central symmetry of the configuration takes place~[3].
If~$\Gamma_1=\Gamma_3$, $\Gamma_2=\Gamma_4$ and vortices form a parallelogram on
a complex plane~$(-z_1=z_3,\ -z_2=z_4)$, then it preserves itself, i.\,e.
in this case there are two invariant relations:
\begin{equation}
\label{e1_23}
M_1=L_1\mbox{, }M_3=L_3.
\end{equation}
Having chosen an origin of coordinates, coinciding with the fixed symmetry
center we can present the equations of motion in the form:
\begin{equation}
\begin{split}
\label{e1_24}
\frac{dM_1}{d\tau}=\Gamma_1M_1L_2(M_3-M_2)+\Gamma_2M_1M_2(L_2-M_3),\\
\frac{dM_3}{d\tau}=\Gamma_1M_3L_2(M_2-M_1)+\Gamma_2M_2M_3(M_1-L_2),\\
\frac{dM_2}{d\tau}=2\Gamma_2M_2L_2(M_1-M_3)\mbox{, }
\frac{dL_2}{d\tau}=2\Gamma_1M_2L_2(M_3-M_1).
\end{split}
\end{equation}

Here the notation: $L_2\equiv M_{24}$ is used and, similarily to (\ref{e1_18}),
the regularization of time~$\tau$:
\begin{equation}
\label{e1_25}
\frac{d\tau}{dt}=\frac{\triangle_4}{2\pi M_1M_2M_3L_2}
\end{equation}
is made. The first pair of equations (\ref{e1_24}) describes motion of the
sides of parallelogram and the second one describes motion of its diagonals.

Equations (\ref{e1_24}) are Hamiltonian with the Hamiltonian:
\begin{equation}
\label{e1_26}
H=2\Gamma_1\Gamma_2\ln M_1+\Gamma_1^2\ln M_2+2\Gamma_1\Gamma_2\ln M_3+\Gamma_2^2\ln L_2.
\end{equation}
The Poisson brackets are quartic by elements of the algebra:
\begin{equation}
\begin{split}
\label{e1_27}
\{M_1, M_2\} & =\frac{1}{\Gamma_1}M_1M_2M_3L_2,\\
\{M_1, M_3\} & =\frac{1}{2}\left(\frac{1}{\Gamma_1}-\frac{1}{\Gamma_2}\right)M_1 M_2 M_3 L_2,\\
\{M_2, M_3\} & =\frac{1}{\Gamma_1}M_1 M_2 M_3 L_2,\\
\{M_1, L_2\} & =-\frac{1}{\Gamma_2}M_1 M_2 M_3 L_2,\\
\{M_2, L_2\} & =0\mbox{, }\{M_3, L_2\}=\frac{1}{\Gamma_2}M_1 M_2 M_3 L_2.
\end{split}
\end{equation}

The Poisson structure can be received from the bracket (\ref{e1_13}) under
the restriction on the mentioned above invariant relations. Now, its rank
is two. The geometric Casimir function is obtained from the relation
between the squares of sides and the squares of diagonals in the parallelogram:
\begin{equation}
\label{e1_28}
F=2(M_1+M_3)-(M_2+L_2).
\end{equation}
For the real motions~$F$ is equal to zero.

The Casimir function (\ref{e1_10}) (``angular momentum'' of the system)
takes the form:
\begin{equation}
\label{e1_29}
D=\Gamma_1M_2+\Gamma_2L_2.
\end{equation}

A reduction of system (\ref{e1_24}) on Casimir functions (\ref{e1_28}),
(\ref{e1_29}) reduces it to the one-dimensional integrable Hamiltonian one~[6].

One must note that in contrast to the rigid body dynamics the equations of
motion in the ``internal'' variables do not have a standard invariant measure
as well as even a measure with the analytical positive density. As a result
dynamics of these problems is deeply different.

A reduction of system (\ref{e1_24}) on Casimir functions (\ref{e1_28}),
(\ref{e1_29}) reduces it to the one-dimensional integrable Hamiltonian one~[6].

One must note that in contrast to the rigid body dynamics the equations of
motion in the ``internal'' variables do not have a standard invariant measure
as well as even a measure with the analytical positive density. As a result
dynamics of these problems is deeply different.

\section{Dynamics of point vortices on a sphere}

\subsection*{Introduction}

Let us consider a problem related to the previous one, a setting of which
also goes back to the last century. In work~[8] I.\,S.\,Gromeka obtained
the equations of motion of point vortices on a sphere. The problem presents
a simple model, describing a motion of vortex formations (cyclones etc.)
in atmosphere of the Earth. The dynamical equations have been written
in~1885~year yet were nearly forgot, though the problem in many respects
is similar to the problem of motion of point vortices on a plane. In the
papers~[9,\,10] there were pointed to integrability of three vortices on
a sphere. In the papers~[11, 12] there were proved the integrability of the
restricted problem of four vortices.

We shall show that the equations of motion of~$N$ vortices on a sphere with
the radius~$R$ can be presented in the Hamiltonian form with degenerated
Poisson structure, assigned by Jacobi algebra~$JQ(C^3_{N+1})$~[13].

In fact, in the usual canonical basis the equations of motion of~$N$ vortices
on a sphere can be written in the form:
\begin{equation}
\label{e2_1}
\dot q_k=\frac{\partial H}{\partial p_k}\mbox{, }\dot p_k=-\frac{\partial H}{\partial q_k},
\end{equation}
where the canonical coordinates~$q_k$, $p_k$ are expressed through the
spherical angles~$\theta_k$, $\varphi_k$ and the strengths~$\Gamma_k$ as
\begin{equation}
\label{e2_2}
q_k=R\sqrt{\Gamma_k}\varphi_k\mbox{, }p_k=R\sqrt{\Gamma_k}\cos\theta_k.
\end{equation}
The Hamiltonian function has the form:
\begin{equation}
\label{e2_3}
H=-\frac{1}{8\pi}\sum_{i, k=1}^N{'}\Gamma_i\Gamma_k\ln(4R^2\sin^2\gamma_{ik}/2),
\end{equation}
where
\begin{equation}
\label{e2_4}
\cos\gamma_{ik}=\cos\theta_i\cos\theta_k+\sin\theta_i\sin\theta_k\cos(\varphi_i-\varphi_k)
\end{equation}
is a cosine of the angle between $i$-th and $k$-th radius vectors, connecting
the center of sphere with the point vortices (see~Fig.~2).

The equations of motion of~$N$ point vortices in the spherical system of
reference have the following form~[9]:
$$
\dot\theta_k=-\frac{1}{4\pi R^2}\sum_{i=1}^N{'}\Gamma_i\frac{\sin\theta_i\sin(\varphi_k-\varphi_i)}
{1-\cos\gamma_{ik}},
$$
\begin{equation}
\label{e2_5}
\sin\theta_k\dot\varphi_k=\frac{1}{4\pi R^2}\sum_{i=1}^N{'}\Gamma_i
\frac{\sin\theta_k\cos\theta_i-\cos\theta_k\sin\theta_i\cos(\varphi_k-\varphi_i)}
{1-\cos\gamma_{ik}}.
\end{equation}
They are Hamiltonian equations with Poisson brackets
\begin{equation}
\label{e2_6}
\{\varphi_i,\,\cos\theta_k\}=\frac{\delta_{ik}}{R^2\Gamma_i}.
\end{equation}

The system (\ref{e2_1}) always has four integrals of motion
$$
H=c_0,
$$
\begin{equation}
\label{e2_7}
F_1\equiv\sum_i\Gamma_i\sin\theta_i\cos\varphi_i=c_1\mbox{, }
F_2\equiv\sum_i\Gamma_i\sin\theta_i\sin\varphi_i=c_2\mbox{, }
F_3\equiv\sum_i\Gamma_i\cos\theta_i=c_3,
\end{equation}
which, however, are not in involution. It is possible to demonstrate that
the integrals~$F_1$, $F_2$, $F_3$ commute as components of angular momentum:
\begin{equation}
\label{e2_8}
\{F_i,\,F_j\}=\varepsilon_{ijk}F_k.
\end{equation}

\subsection*{Algebraic representation}

Let us introduce the ``internal'' variables:
\begin{equation}
\label{e2_9}
M_{ij}\equiv4R^2\sin^2\gamma_{ij}/2,
\end{equation}
being the squares of spans between arcs of corresponding vortices. The
Hamiltonian (\ref{e2_3}) depends only on introduced mutual variables:
\begin{equation}
\label{e2_10}
H=-\frac{1}{8\pi}\sum_{i,\,k=1}^N{'}\Gamma_i\Gamma_k\ln M_{ik}.
\end{equation}

From the relations between the canonical coordinates~$q_i$, $p_j$ (\ref{e2_6})
it is possible to find commutators between the values~$M_{ik}$:
\begin{equation}
\label{e2_11}
\{M_{ij},\,M_{kl}\}=\frac{4}R\left(\frac{1}{\Gamma_i}\delta_{ik}-\frac{1}{\Gamma_j}\delta_{jk}
\right)\triangle_{ijl}+\frac{4}{R}\left(\frac{1}{\Gamma_i}\delta_{il}-\frac{1}{\Gamma_j}\delta_{jl}
\right)\triangle_{ijk},
\end{equation}
where the following notations are used:
\begin{equation}
\label{e2_12}
\triangle_{ijk}\equiv{\bf r}_i\wedge{\bf r}_j\wedge{\bf r}_k.
\end{equation}
Let us note that we use the same notations in both considered problems. Thus,
the Poisson brackets between the values~$M_{ik}$ are proportional to external
form of volume of parallelepiped spanned on the radius vectors of triple of
vortices on a sphere.

It is possible to demonstrate that the algebra of variables~$M_{ik}$
and~$\triangle_{ijk}$ is closed.
\begin{equation}
\begin{split}
\label{e2_13}
\{M_{ij},\,\triangle_{klm}\} =
R\left(\frac{1}{\Gamma_i}\delta_{ik}-\frac{1}{\Gamma_j}\delta_{jk}\right)
(M_{li}-M_{im}+M_{mj}-M_{jl})+\\
+ R\left(\frac{1}{\Gamma_i}\delta_{il}-\frac{1}{\Gamma_j}\delta_{jl}\right)
(M_{mi}-M_{ik}+M_{kj}-M_{jm})+\\
+ R\left(\frac{1}{\Gamma_i}\delta_{im}-\frac{1}{\Gamma_j}\delta_{jm}\right)
(M_{ki}-M_{il}+M_{lj}-M_{jk})+\\
+ \frac{1}{2R}\left(\frac{1}{\Gamma_i}\delta_{ik}-\frac{1}{\Gamma_j}\delta_{jk}\right)
(M_{jl}M_{im}-M_{mj}M_{il})+\\
+ \frac{1}{2R}\left(\frac{1}{\Gamma_i}\delta_{il}-\frac{1}{\Gamma_j}\delta_{jl}\right)
(M_{jm}M_{ik}-M_{jk}M_{im})+\\
+ \frac{1}{2R}\left(\frac{1}{\Gamma_i}\delta_{im}-\frac{1}{\Gamma_j}\delta_{jm}\right)
(M_{jk}M_{il}-M_{ik}M_{jl});
\end{split}
\end{equation}

\begin{equation}
\begin{split}
\label{e2_14}
\{\triangle_{ijk},\,\triangle_{lmn}\}= &
\frac{\delta_{il}}{\Gamma_i}\left(R(\triangle_{jkn}-\triangle_{jkm})+
\frac{1}{2R}(M_{in}\triangle_{jkm}-M_{im}\triangle_{jkn})\right)+\\
+ & \frac{\delta_{im}}{\Gamma_i}\left(R(\triangle_{jkl}-\triangle_{jkn})+
\frac{1}{2R}(M_{il}\triangle_{jkn}-M_{in}\triangle_{jkl})\right)+\\
+ & \frac{\delta_{in}}{\Gamma_i}\left(R(\triangle_{jkm}-\triangle_{jkl})+
\frac{1}{2R}(M_{im}\triangle_{jkl}-M_{il}\triangle_{jkm})\right)+\\
+ & \frac{\delta_{jl}}{\Gamma_j}\left(R(\triangle_{ikm}-\triangle_{ikn})+
\frac{1}{2R}(M_{jm}\triangle_{ikn}-M_{jn}\triangle_{ikm})\right)+\\
+ & \frac{\delta_{jm}}{\Gamma_j}\left(R(\triangle_{ikn}-\triangle_{ikl})+
\frac{1}{2R}(M_{jn}\triangle_{ikl}-M_{jl}\triangle_{ikn})\right)+\\
+ & \frac{\delta_{jn}}{\Gamma_j}\left(R(\triangle_{ikl}-\triangle_{ikm})+
\frac{1}{2R}(M_{jl}\triangle_{ikm}-M_{jm}\triangle_{ikl})\right)+\\
+ & \frac{\delta_{kl}}{\Gamma_k}\left(R(\triangle_{ijn}-\triangle_{ijm})+
\frac{1}{2R}(M_{kn}\triangle_{ijm}-M_{km}\triangle_{ijn})\right)+\\
+ & \frac{\delta_{km}}{\Gamma_k}\left(R(\triangle_{ijl}-\triangle_{ijn})+
\frac{1}{2R}(M_{kl}\triangle_{ijn}-M_{kn}\triangle_{ijl})\right)+\\
+ & \frac{\delta_{kn}}{\Gamma_k}\left(R(\triangle_{ijm}-\triangle_{ijl})+
\frac{1}{2R}(M_{km}\triangle_{ijl}-M_{kl}\triangle_{ijm})\right).
\end{split}
\end{equation}

The quadratic algebra of vortices on a sphere is presented as a deformation
of the linear algebra of vortices on a plane using the parameter~--- the
radius of curvature of sphere~$R$. The algebra of vortices on a plane
(\ref{e1_7})--(\ref{e1_9}) and the corresponding Casimir functions of system
on a sphere are obtained in the limit~$R\to\infty:$ $\triangle^v\bigm/R\to\triangle$,
$M^v\to M$ from the algebra (\ref{e2_11})--(\ref{e2_14}). Here we symbolically
marked by a symbol~$v$ ihe variables of the problem on the sphere.

The linear Casimir functions of the quadratic algebra analogous to the
corresponding functions of the linear algebra (\ref{e1_10})--(\ref{e1_11}).
The analog of the relations (\ref{e1_12}) can be written as follows:
\begin{equation}
\begin{split}
\label{e2_15}
F_{ijk}= & 2\triangle_{ijk}{}^2+R^2(M_{ij}^2+M_{ik}^2+M_{jk}^2)-\\
- & 2R^2(M_{ij}M_{jk}+M_{ik}M_{jk}+M_{ij}M_{ik})+M_{ij}M_{jk}M_{ki}=0.
\end{split}
\end{equation}

\subsection*{Partial cases of integrabilily of three and four vortices
on a sphere}

In the spherical problem, the analogue of the partial case on the plane is
a situation when the integrals of motion (\ref{e2_7}) have zero values:
\begin{equation}
\label{e2_16}
F_1=F_2=F_3=0.
\end{equation}
The case is completely integrable because there are sufficient number of
commuting, on zero level, integrals.

It is not difficult to obtain identity, analogous to (\ref{e1_21}):
\begin{equation}
\label{e2_17}
D/2=\left(R\sum_{i=1}^N\Gamma_i\right)^2-F_1^2-F_2^2-F_3^2.
\end{equation}
Consequently, the considered case corresponds to the level of Casimir
function:
$$
D/2=\left(R\sum_{i=1}^N\Gamma_i\right)^2.
$$
In order to find the invariant relations in variables~$M_{ij}$ it is possible
to use identity, between absolute and mutual coordinates, analogous
to (\ref{e1_22}):
\begin{equation}
\label{e2_18}
f_{ij}=\frac{1}{2}\sum_{k=1}^N\Gamma_k(M_{jk}-M_{ik})=
R(F_1(x_i-x_j)+F_2(y_i-y_j)+F_3(z_i-z_j)).
\end{equation}

\section{Stationary vortex states}

The obtained forms of the dynamical equations on a plane and on a sphere can be
successfully used for finding the steady-state configurations, being the
partial solutions of the equations of motion.

Requirements of~$N(N-1)/2$ mutual distances~$\dot M_{ij}=0$ being constant are
the conditions of stationary state of vortex configurations. They have the same
form both for the configurations on a plane and on a sphere in the ``internal''
variables as its Hamiltonians (\ref{e1_2}), (\ref{e2_10}) and commutation
relations~$\{M_{ij},\,M_{kl}\}$ (\ref{e1_7}), (\ref{e2_10}) have the same form:
\begin{equation}
\label{e3_1}
\sum_{l=1}^N{'}\left(\frac{1}{M_{il}}-\frac{1}{M_{jl}}\right)\Gamma_l\triangle_{ijl}(M)=0.
\end{equation}

The conditions (\ref{e3_1}) are very suitable for finding symmetric
configurations. On the plane there is solution of~$N$ vortices of equal
strengths~$\Gamma$, disposed in vertices of regular polygon, refined into a
circle of radius~$R_0$. The system rotates with angular velocity
\begin{equation}
\label{e3_2}
\Omega=\frac{\Gamma(N-1)}{4\pi R_0^2}.
\end{equation}
The analysis of stability of such configurations in linear approximation was
carried out in~[15,\,16].

For investigation of such configurations J.\,J.\,Thomson was awarded with
Adams~prize in~1883~[15]. These solutions were at the basis of the theory
of vortex atoms propagandized by W.\,Kelvin (before quantum mechanics).
Analogous configuration of identical vortices on a sphere is disposed,
according to (\ref{e3_5}), on latitude~$\theta=\theta_0$, coordinates~$\varphi$ are
connected by conditions: $\varphi_k-\varphi_i=(k-i)2\pi\bigm/N$ (see~Fig.~3).
Angular velocity of rotation about the axis~$z$ is:
\begin{equation}
\label{e3_3}
\Omega=\frac{\Gamma(N-1)}{4\pi R_0^2}\cos\theta_0.
\end{equation}

The angular velocity of the chain of vortices decreases from poles to equator
and on the equator the configuration is static.

The next solution of the system (\ref{e3_1}) are collinear configurations:
$N$~identical vortices on a line rotate with some angular velocity around axis
perpendicular to the plane. On the sphere of radius~$R$ the collinear
configuration ( $N$ atoms on the meridian) rotates around the axis~$z$ with
angular velocity~$\Omega$ (see~Fig.~4).

Collinear configurations of vortices on a sphere were not discussed in
literature before. We can find them using the following conditions:
coordinates~$\varphi_i$ of the point vortices have the equal values or differ
on~$\pi$. In addition: $\triangle_{ijl}({\bf M})=0$. It is easy to notice that
coordinates~$\theta_i$, $(0\leqslant\theta\leqslant 2\pi)$ are roots of the following system
of trigonometric equations:
\begin{equation}
\label{e3_4} \frac{4\pi
R^2\Omega}\Gamma\sin\theta_k=\sum_{i=1}^N{'}\cot\left(\dfrac{\theta_k-\theta_i}2\right)\mbox{,
}(1\leqslant k\leqslant N).
\end{equation}

The chain of vortices (\ref{e3_4}) deserves more attention. The roots of the
system assigned the collinear configuration~$\theta_i$ can be found as equilibrium
position of the system of~$N$ particles on a circle set by the Hamiltonian
\begin{equation}
\label{e3_5}
H=\frac{1}{2}\sum_{k=1}^Np_k^2+\frac{4\pi R^2\Omega}\Gamma\sum_{k=1}^N\cos\theta_k+
\sum_{i,\,k=1}^N{'}\ln\left|\sin\frac{\theta_k-\theta_i}2\right|.
\end{equation}
Evidently, the positions of equilibrium of the chain of particles coincide with
roots of the system (\ref{e3_4}). Such connection in plane case was marked by
Calogero~[17]. The locations of identical vortices on a line are set by zeros
of Hermitean polynomials of~$N$-th power. The system (\ref{e3_5}) under~$\Omega=0$
was considered in~[18], where statistical properties of levels of energies of
one-dimensional classical Coulomb gas are considered. In the paper~[19] the
results of equilibrium analysis of the system (\ref{e3_5}) under
condition~$\Omega=0:$ $\theta_i=\theta_0+k\pi\bigm/N$, $k=1,\,\ldots\,,\,N$ are adduced.

The problem defined by the Hamiltonian (\ref{e3_5}) is usually nonintegrable.

\subsection*{Acknowledgements}

The work is supported partially by the grant of RFBR (grant No. 96--01--00747).

\begin{figure}[ht!]
$$
\includegraphics{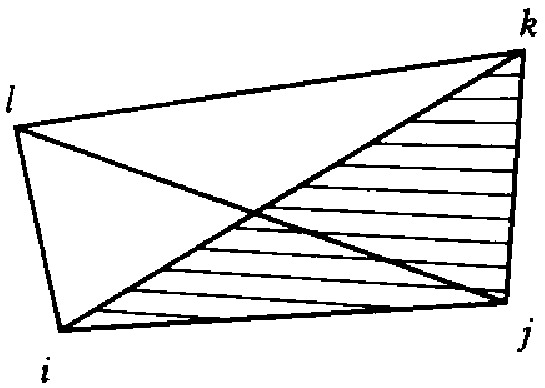}
$$
\caption{}
\end{figure}

\begin{figure}[ht!]
$$
\includegraphics{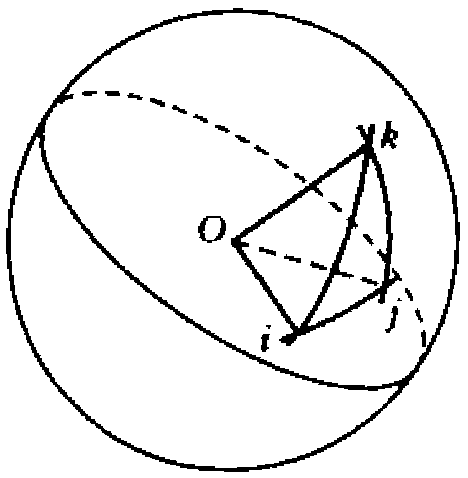}
$$
\caption{}
\end{figure}

\begin{figure}[ht!]
$$
\includegraphics{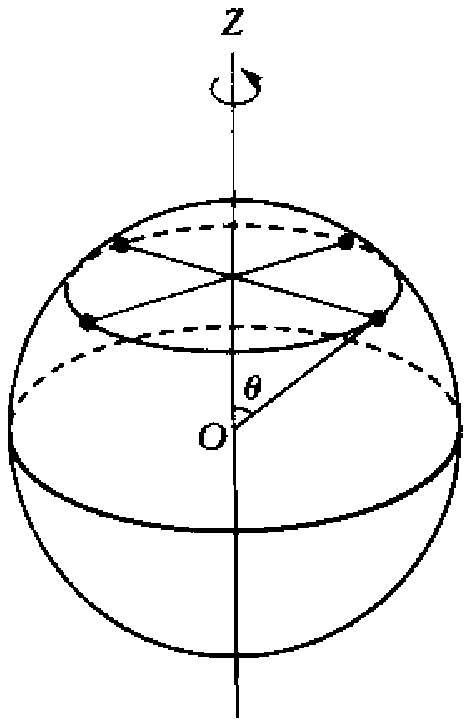}
$$
\caption{}
\end{figure}

\begin{figure}[ht!]
$$
\includegraphics{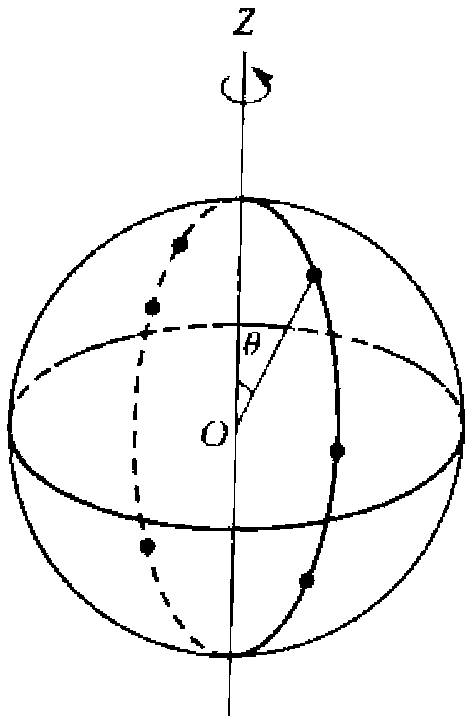}
$$
\caption{}
\end{figure}

\end{document}